\documentclass[review]{elsarticle}

\usepackage{lineno,hyperref}
\usepackage{amsmath}
\usepackage{graphicx}
\usepackage{hyperref}
\usepackage{xcolor}
\usepackage{multirow}
\usepackage{makecell}
\modulolinenumbers[5]

\journal{Journal of \LaTeX\ Templates}

\begin{document}
	
	\begin{frontmatter}
		
		\title{Mechanical properties of 2D metal-organic and covalent-organic frameworks with non trivial topological band dispersion}
		
		\author{Priyadarshini Kapri, Takuto Kawakami, and Mikito Koshino}
		
		\address{Department of Physics, Osaka University, Osaka 560-0043, Japan}
		
		
		\ead{priya.kapri@qp.phys.sci.osaka-u.ac.jp, t.kawakami@qp.phys.sci.osaka-u.ac.jp, koshino phys.sci.osaka-u.ac.jp}







\begin{abstract}
Using density functional theory (DFT), we investigate mechanical properties of a few 2D metal-organic frameworks (MOFs) and covalent-organic frameworks (COFs) having Dirac and flat bands. These porous materials have become a subject of great captivation because of their physical stability, distinctive structural characteristics and large surface to volume ratio.  The inherent porosity of these frameworks gives rise to many fascinating and occasionally surprising phenomena, which makes them potential candidates for technological applications.  For reliable usage of MOFs/COFs in functional nanodevice and practical application, it is quite imperative to investigate their mechanical properties.  
Thus, herein a particular attention is paid to study elastic deformation of few 2D MOFs and COFs having non trivial topological band dispersion in the regime  with linear dependency of stress upon strain. Specially, we consider different types of deformation and find all the components of  elastic tensor from the stress-strain  and energy-strain curves. These findings may provide useful information to fabricate the MOFs/COFs based devices by lowering the number of experiments. 
\end{abstract}


\end{frontmatter}


\section{Introduction}
The striking realization of graphene \cite{Nov1,ADM} has advanced an emerging field of 2D materials including the 2D topological and quantum materials \cite{CLK,BAB}. Graphene, a two dimensional single layer of hexagonal lattice of carbon atoms, holds an exotic low energy excitations which are linear in momentum ($E({\bf k})\propto {\bf k}$). Owing to the completely planar atomic nature, graphene and its derivatives, such as graphene thin films, graphene nanoflakes have garnered an enormous impetus in the field of low dimension and surface physics. Along with this exotic linear band dispersion, recently, realization of a completely non dispersive flat band ($E({\bf k})\propto {\bf k^0}$) in certain geometrically frustrated lattices with special symmetry, such as 2D Kagome (corner shared triangle) and Leib (edge centered square) lattices \cite{HT,AM,SM,YZ1} has been gaining  the attention of community. In contrast to the Dirac dispersion, which features mass less electrons, this non trivial topological flat band exhibits electrons with an infinite mass.  The topological flat band arises because of the phase cancellation of Bloch wave functions (destructive quantum interference) \cite{ZL1} and accompanies a localized state by canceling outward lattice hopping.
Topological materials with non trivial band topology like Dirac and non dispersive flat bands have promising applications for fabricating devices in the field of quantum information, computation as well as in spintronics. 

Recently, the 2D organic topological and quantum materials, specially metal-organic framework (MOF,  crystalline porous organic-inorganic hybrid materials with a regular arrangement of metal ions encircled by organic `linker' molecules) and covalent organic framework (COF, a group of crystalline porous organic polymers having highly organized structures and persistent porosity) with Dirac and flat bands \cite{ZFW1,WJ,AMS,ZMH,XLQ} have been gaining interest for the fundamental interest embedded therein and possible potential applications. The main advantages of MOFs/COFs are their unique structural properties, i.e., high surface to volume ratio and large porosity.  Besides, the vast and remarkable diversity of co-ordination chemistry provides numerous combinations of metal ions and organic linker and hence offers a huge range of lattice, orbital and spin symmetries \cite{XF,HCJZ} and eventually makes the MOFs/COFs an ideal platform for hosting the exotic Dirac and non dispersive flat bands. Lately, the experimental advancements in manufacturing $\pi$ conjugated MOFs/COFs with robust magnetization and high conductivity yield additional impetus.  These MOFs/COFs host many exotic quantum states which were realized previously only in inorganic systems. In an inorganic crystal, generally all the atomic orbitals belong to the one common group of lattice symmetry, whereas for organic systems, atomic orbitals for metal ion and molecular orbitals for different organic ligands are subject to different sub groups of lattice symmetry. Thus, the band structures of MOFs/COFs are more complicated and lead to more exotic topological quantum states. The diversity and high tunability makes the MOFs/COFs a great playground  for exploring novel quantum physics, quantum chemistry and promising applications.

For practical application and reliable usage of MOFs/COFs in fabrication of nano-devices, it is mandatory to investigate their energetics as well as mechanical properties. The mechanical failure or undesirable deformation can leads to the overall failure of the device. Though graphene based nanoribbons (building blocks \cite{AC,BR} of many nano-devices, such as resonators, nano-switches, where sensitivity of the resonator and the pull-in voltage of nano-switches depend on the stiffness constant of the graphene cantilever beam \cite{PL}) have been extensively studied \cite{HB,HZ,CR,RF,AVO}, the mechanical properties of MOFs/COFs have been somehow overlooked.  Understanding the mechanical properties of MOFs/COFs  plays an crucial role in the post-synthetic processing, and quickens the successful transition from academic interest to industrial relevance for fabricating MOFs/COFs based devices by reducing the number of experiments.

Motivated by the above discussions, in this paper, we  study the mechanical properties of a few 2D MOF and COF structures with non trivial band dispersion. 
The paper is organized as follows. Section (\ref{Sec2}) includes the basic theory to calculate the different elastic constants along with the computational methodology. Section (\ref{Sec3}) presents the numerical results and the corresponding discussions. Finally, Sec. (\ref{Sec4}) concludes our main findings.
\section{Theory and Methodology}
\label{Sec2}

\begin{figure}
	\centering
	\includegraphics[width=0.70\textwidth]{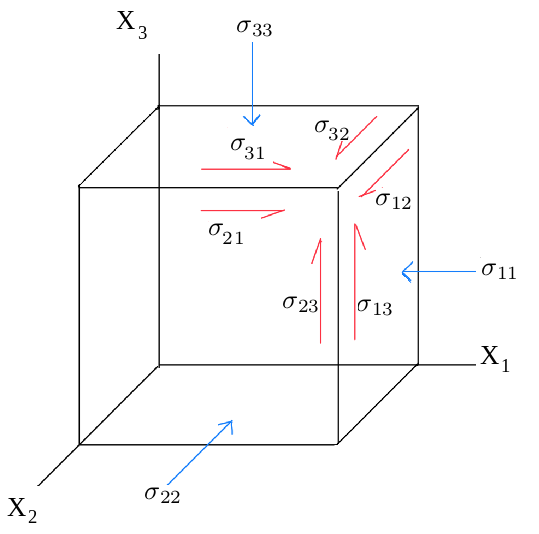}
	\caption{A schematic illustration of a 3D structure with forces acting on it in three dimensions.}
	\label{FigSS}
\end{figure}
In this section, we revisit the fundamentals of elastic theory to study the simple deformations to determine the corresponding elastic constants. Initially, we start with a generic 3D system, however, as the main focus of this paper is the systems confined in a plane, we elaborate our discussions considering the plane deformation.

Elastic deformation of any material can be described by the familiar Hooke's law, where  the proportionality of stress and strain plays the key role.  Stress is defined as the force per unit area within the materials that arises due to the externally applied forces or any deformation, while strain is the relative change in the positions of points within a body that has undergone any deformation. Beyond the elastic domain, higher order effects become dominant and the material usually gets damaged.
For an external arbitrary load, one can measure the stress on it in various directions (see Fig. (\ref{FigSS})), which makes the stress a second rank tensor
$\sigma_{ij}$ with $i=1,2,3$ and $j=1,2,3$. The stress tensor is defined as $\sigma_{ij}=\sum_k\sum_l\gamma_{ijkl}\epsilon_{kl}$ with $\gamma_{ijkl}$ and $\epsilon_{kl}$ ($k=1,2,3$ and $l=1,2,3$) being the elastic and strain tensors of rank $4$  and rank $2$, respectively. However, because of the symmetric nature of the stress, strain and elastic tensors, only $21$ out of $81$ components of the elastic tensor are independent. Further, more additional symmetries, such as crystal symmetry reduce the number of independent components.

To avoid the cumbersome notations of elastic tensor, a simple notation named as `Voigt notation' is generally preferred, where rank $2$ tensors are rewritten as vectors and the elasticity tensor is written as a $6\times6$ matrix. Thus, the Voigt notation allows to reconstruct the Hooke's law in a much more compact format without any drawbacks. However, once the Voigt notation is applied, one cannot apply any co-ordinate transformation anymore. The usual convention in Voigt notation to rewrite the stress and strain tensors are as follows
\begin{equation}
	\begin{pmatrix}
		\sigma_{11} & \sigma_{12} &\sigma_{13}\\
		\sigma_{12} &\sigma_{22} & \sigma_{23}\\
		\sigma_{13}& \sigma_{23} & \sigma_{33}
	\end{pmatrix}\rightarrow \begin{pmatrix}
		\sigma_{11} \\
		\sigma_{22} \\
		\sigma_{33}\\
		\sigma_{23} \\
		\sigma_{13} \\
		\sigma_{12}
	\end{pmatrix}\rightarrow \begin{pmatrix}
		\sigma_{1} \\
		\sigma_{2} \\
		\sigma_{3}\\
		\sigma_{4} \\
		\sigma_{5} \\
		\sigma_{6}
	\end{pmatrix},
\end{equation}
and 
\begin{equation}
	\begin{pmatrix}
		\epsilon_{11} & \epsilon_{12} & \epsilon_{13}\\
		\epsilon_{12} &\epsilon_{22} & \epsilon_{23}\\
		\epsilon_{13}& \epsilon_{23} & \epsilon_{33}
	\end{pmatrix}\rightarrow \begin{pmatrix}
		\epsilon_{11} \\
		\epsilon_{22} \\
		\epsilon_{33}\\
		2\times\epsilon_{23} \\
		2\times\epsilon_{13} \\
		2\times\epsilon_{12}
	\end{pmatrix}\rightarrow \begin{pmatrix}
		\epsilon_{1} \\
		\epsilon_{2} \\
		\epsilon_{3}\\
		\epsilon_{4} \\
		\epsilon_{5} \\
		\epsilon_{6}
	\end{pmatrix}.
\end{equation}
Similarly, for the elastic tensor, the tensor notation: [$11$; $22$; $33$; $23,32$; $31,13$; $12,21$] is replaced by the matrix notation: [$1$; $2$; $3$; $4$; $5$; $6$].

For a 2D structure, confined in a plane, we can omit $\epsilon_{3}$, $\epsilon_{4}$ and  $\epsilon_{5}$ from the strain vectors. Hence, with the plane deformation, for a 2D structure, the stress-strain relation can be written as  
\begin{equation}
	\label{Eq3}
	\begin{pmatrix}
		\sigma_{1} \\
		\sigma_{2} \\
		\sigma_{6}\end{pmatrix}	= \begin{pmatrix}
		\gamma_{11} & \gamma_{12} & \gamma_{16}\\
		\gamma_{12} &\gamma_{22} & \gamma_{26}\\
		\gamma_{16}& \gamma_{26} & \gamma_{66}
	\end{pmatrix}	\begin{pmatrix}\epsilon_{1} \\
		\epsilon_{2} \\
		\epsilon_{6}\end{pmatrix}.
\end{equation}
Thus, for a 2D structure the elastic matrix is now reduced to a $3\times3$ matrix, where the independent components are only $6$. However, as mentioned earlier, the lattice symmetry further reduces the actual number of the independent components. All the structures of our consideration have hexagonal symmetry and hence the number of independent components should be only $2$, as $\gamma_{11}=\gamma_{22}$, $\gamma_{16}=\gamma_{26}=0$, and $\gamma_{66}=\frac{1}{2}(\gamma_{11}-\gamma_{12})$ \cite{JFN}.

To determine the components of the elastic tensor, one needs to apply some small but finite deformations and obtain the corresponding energy or stress acting on the unit cell.  For in-plane deformation, the deformation tensor can be described by a symmetric $2\times 2$ matrix:
\begin{equation}\delta=
	\begin{pmatrix}
		1+\epsilon_{1} & \epsilon_{12} \\
		\epsilon_{12} &1+\epsilon_{2} 
	\end{pmatrix},
\end{equation}
which deforms the unit cell vectors via ${\bf v}_{i}^{\prime}=\delta {\bf v}_{i}$ with 
${\bf v}_{i}^{\prime}$ and  ${\bf v}_{i}$ being the unit cell vectors after and before the deformation, respectively.
The elastic constants can be obtained from the stress-strain curve (linear fitting) as well from the energy-strain curve (quadratic fitting).
We only require two different types of deformations to evaluate our systems because they should have only two independent components (hexagonal symmetry). However, we calculate each component separately and verify the relationships between them. Hence, in order to compute all the elastic constants, we take into account four different forms of deformation as follows:
stretching along (i) $\hat{x}$ axis ($\epsilon_1=\eta$, $\epsilon_{12}=\epsilon_{2}=0$) and (ii) $\hat{y}$  ($\epsilon_2=\eta$, $\epsilon_{12}=\epsilon_{1}=0$) axis allow to calculate $\gamma_{11}$ and $\gamma_{22}$, via $\gamma_{11}=\frac{\sigma_1}{\eta}$ and  $\gamma_{22}=\frac{\sigma_2}{\eta}$ (see Eq. ({\ref{Eq3}})), respectively. Similarly, stretching along (iii) $\hat{x}$ and $\hat{y}$ axes simultaneously ($\epsilon_{1}=\epsilon_{2}=\eta$, $\epsilon_{12}=0$) yields the value of $\gamma_{12}$ from the relations: $\gamma_{12}=\frac{\sigma_1}{\eta}-\gamma_{11}$ and $\gamma_{12}=\frac{\sigma_2}{\eta}-\gamma_{22}$. Finally, (iv) the shearing ($\epsilon_{12}=\eta$, $\epsilon_{1}=\epsilon_{2}=0$) allows to calculate $\gamma_{16}=\frac{\sigma_1}{2\eta}$, $\gamma_{26}=\frac{\sigma_2}{2\eta}$ and 
$\gamma_{66}=\frac{\sigma_6}{2\eta}$. Thus, simply applying four types of deformation, one can obtain all the elastic constants using aforementioned stress-strain relations for a structure confined in a plane. The stress evaluation method confirms that  $\gamma_{16}$ and  $\gamma_{26}$ are zero for all of our systems having hexagonal symmetry.

As mentioned earlier, the energy-strain curve also yields the different elastic constants.
The energy associated with an elastic deformation can be expressed using
equation \cite{MN}
\begin{equation}
	E=E_0+A_0\sum_i\sum_j\sigma_{ij}\epsilon_{ij}+\frac{A_0}{2}\sum_{i}\sum_{j}\epsilon_{ij}\cdot(\sum_k\sum_l\gamma_{ijkl}\epsilon_{kl}),
\end{equation}
with $A_0$ being the area of un-deformed unit cell.
Applying the four deformations discussed earlier
gives rise to four different forms of this equation
\begin{eqnarray}
	E_{\eta}&=&E_0+A_0\sigma_1\eta+\frac{A_0}{2}\gamma_{11}\eta^2,\\\nonumber
	E_{\eta}&=&E_0+A_0\sigma_2\eta+\frac{A_0}{2}\gamma_{22}\eta^2,\\\nonumber
	E_{\eta}&=&E_0+A_0(\sigma_1+\sigma_2)\eta+\frac{A_0}{2}(\gamma_{11}+\gamma_{11}+2\gamma_{12})\eta^2,\\\nonumber
	E_{\eta}&=&E_0+2A_0\sigma_6\eta+2A_0\gamma_{66}\eta^2,
\end{eqnarray}
which allow to evaluate the values of $\gamma_{11}$, $\gamma_{22}$, $\gamma_{12}$ and $\gamma_{66}$, respectively from the quadratic fitting of the total energy vs. strain curve.   

In order to perform the DFT theoretical simulations, the Quantum-Espresso computer code \cite{PG1,PG2} with an ultra-soft RRKJ-type pseudopotential has been used. 
The Quantum-Espresso package allows to calculate the total energy and the stress tensor of any
periodic arrangement of atoms.  Initially, we optimize the structure as far as the atomic positions and the unit-cell
are concerned. Subsequently, we deform the structure with strain by
increasing or decreasing the unit-cell dimension, and keeping the dimension fixed in the calculations with all the atomic
positions being allowed to relax. It is to be noted that the Quantum-Espresso package
can only calculate three dimensional materials, and thus to
treat a 2D layer (confined in $x-y$ plane), one needs to insert a large enough vacuum gap  in
the $\hat{z}$-direction to neglect the interaction between the individual layers. We consider an empty space of at least 15 {\AA} 
 between the two layers, however after the full relaxation that value slightly changes. In order to convert the 3D stress (force per area) obtained in Quantum-Espresso DFT calculation into 2D stress (force per length), one needs to multiply the 3D value by inter-planar spacing.

\section{Results and Discussions}
\label{Sec3}
This section presents the numerical results of elastic constants of a few metal-organic and covalent-organic frameworks. In the initial part, we demonstrate three metal organic frameworks, where we basically consider the structures: 2D indium-phenylene organometallic framework (In$_2$(C$_6$H$_4$)$_3$), Ni$_3$(C$_{6}$S$_{6}$)$_2$ and Cu-hexaiminobenzene framework [Cu$_3$(HAB)$_2$, Cu$_3$(C$_6$N$_6$H$_6$)$_2$], which are all known to host the flat bands. The later part presents the covalent-organic frameworks. Here, we take into account two graphene nanoflakes with periodic H passivated regular  nano-holes with different sizes and a covalent carbon nitride network with a C$_9$N$_4$ stoichiometry.  All the elastic constants obtained for the MOF and COF structures are tabulated in two tables.
\begin{figure}[h!]
	\centering
	\includegraphics[width=0.85\textwidth]{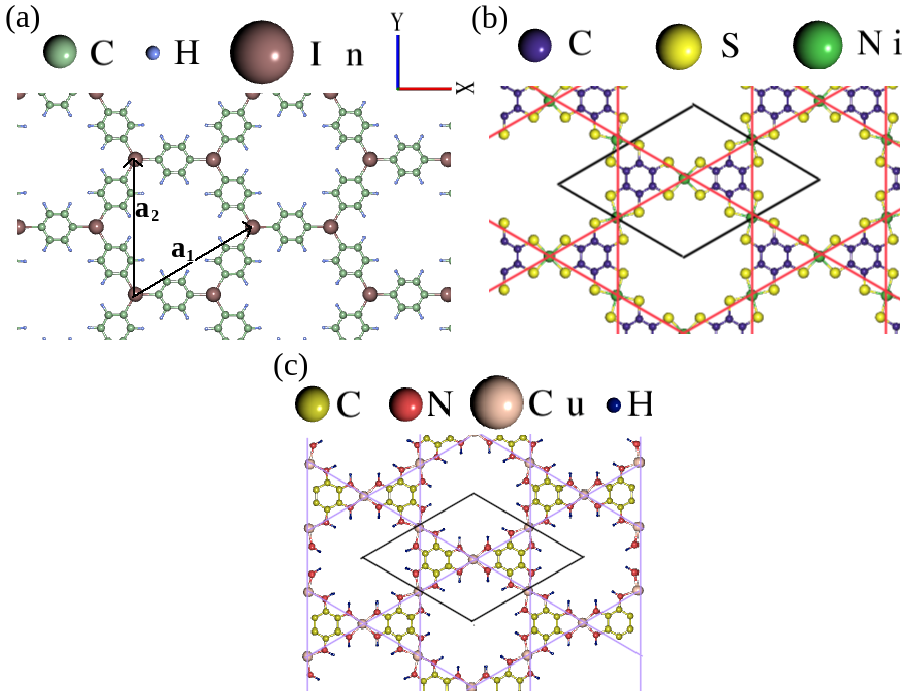}
	\caption{The atomic structures of (a) the IPOF (In$_2$(C$_6$H$_4$)$_3$) depicting  the binding of In ($p$-orbital heavy elements) with organic ligands (paraphenylenes) into a hexagonal lattice structure, (b) the Ni$_3$(C$_6$S$_6$)$_2$ with black line showing the unit cell and the red lines framing the Kagome lattice, and (c) the Cu$_3$(C$_6$ N$_6$ H$_6$)$_2$, where
		the black line indicates the unit cell.  The Cu ions form a Kagome lattice
		connected by purple lines. }
	\label{FigMOF}
\end{figure}
\subsection {MOF 1: In$_2$(C$_6$H$_4$)$_3$}
\begin{table}[h!]
	\centering
	\begin{tabular}{| *{9}{c|} }
		\hline
		& \multicolumn{2}{c|}{MOF 1}
		& \multicolumn{2}{c|}{MOF 2}
		& \multicolumn{2}{c|}{MOF 3}
		\\
		\hline
		\thead {Elastic \\ Constants}  &   \thead {S-S\\(eV/{\AA}$^2$)}  &   \thead {TE-S\\ (eV/{\AA}$^2$)}  &    \thead {S-S\\(eV/{\AA}$^2$)}  &   \thead {TE-S\\ (eV/{\AA}$^2$)}  &    \thead {S-S\\(eV/{\AA}$^2$)}  &   \thead {TE-S\\ (eV/{\AA}$^2$)}    \\
		\hline
		$\gamma_{11}$   &   2.05  &   1.94  &   2.30  &   2.34  &   1.51  &   1.52   \\
		\hline
		$\gamma_{22}$   &   1.91  &   1.82  &   2.35  &   2.34  &   1.65  &   1.55    \\
		\hline
		$\gamma_{12}$   &   0.91  &   0.99  &   0.95  &   0.93  &   0.46  &   0.35    \\
		\hline
		$\gamma_{66}$   &   0.49  &   0.55  &   0.69  &   0.74  &   0.57  &   0.59   \\
		\hline
	\end{tabular}\\
	\caption{Elastic constants of metal-organic frameworks, where S-S  and TE-S stand for the calculations from stress-strain and total energy-strain curves, respectively.} 
	\label{tab:MOF} 
\end{table}
Initially we consider a 2D indium-phenylene organometallic framework (IPOF). This unique IPOF structure
exhibits a nearly flat band 
around the Fermi level characterized by a nontrivial $Z_2$ topological number \cite{ZL2}.  The atomic structure of the IPOF (In$_2$(C$_6$H$_4$)$_3$) is shown in Fig. (\ref{FigMOF}a), which manifests the binding of In ($p$-orbital heavy elements) with organic ligands (paraphenylenes) into a hexagonal lattice structure. The In atoms intrinsically bonds to three phenylenes
in a planar triangular geometry, which is a common feature of group-III
elements. Thus, each hexagonal unit cell of In$_2$(C$_6$H$_4$)$_3$ contains two In atoms and three phenylenes. 

For DFT calculation, a Monkhorst-Pack grid of $8 \times 8 \times 1$ k-points
are taken for reciprocal integrations of the structure. Further, valence wave functions and the deficit charge density are
expanded in terms of the plane-wave basis set with cutoff
energies of $80$ Ry and $480$ Ry, respectively.
As mentioned earlier, the structure is confined in $xy$ plane, thus, the spurious interaction
between images in the perpendicular direction 
is avoided by keeping an empty space of at least $15$ {\AA} in the
$\hat{z}$-directions.   We perform the full optimization of the atomic structures until the force acting on each atom
is less than $0.001$ Ry/Bohr. Here, it is to be noted that the unrelaxed lattice constant and the relaxed lattice constants are different and the vacuum space in the $\hat{z}$ direction slightly changes after the full optimization of the atomic structures. By minimizing the total energy as a function of
lattice constants, the actual lattice constant is obtained. The optimized lattice constant is found to have the value $14.85$ $\AA$. Strains are applied in
the increments of $0.001$ and the atomic structure is geometrically
optimized to find the energy and stress at that value of strain. The atomic coordinates of the last step is used as
the initial guess for the next step to fasten the
convergence. 

As mentioned earlier, for calculating the elastic constants, we  consider four types of deformation. The plots for the variations of 3D stress and total energy as a function of strain $\eta$ for all the deformations are illustrated in the Appendix. With type $(i)$ deformation (stretching along $\hat{x}$ axis), the stress and energy evaluation methods yield the values of $\gamma_{11}=2.05$ eV/{\AA}$^2$ and $1.94$ eV/{\AA}$^2$, respectively (see Figs. (\ref{FigMOF1CC}a and (\ref{FigMOF1CC}b). Thus, the results obtained from stress and energy evaluation methods deviate by $0.11$ eV/{\AA}$^2$.  Similarly, with type $(ii)$ deformation, i.e., stretching along $\hat{y}$ axis, we obtain the values of $\gamma_{22}=1.91$  eV/{\AA}$^2$ and  $1.82$ eV/{\AA}$^2$ (see Figs. (\ref{FigMOF1CC}c) and (\ref{FigMOF1CC}d)), where also results for the two methods deviate by $0.11$  eV/{\AA}$^2$. Despite the fact that $\gamma_{11}$ and $\gamma_{22}$ should have the same values (because of the hexagonal symmetry), they differ by $0.14$  eV/{\AA}$^2$ for the stress evaluation technique and $0.12$ eV/{\AA}$^2$ for the total energy evaluation method, which  are reliable differences indicating just a numerical error. Next, with type $(iii)$ deformation (stretching along both the $\hat{x}$ and $\hat{y}$ axes), the stress evaluation method yields $\gamma_{12}=0.91$ eV/{\AA}$^2$ (using $\gamma_{12}=\frac{\sigma_{22}}{\eta}-\gamma_{11}$), whereas the total energy evaluation method provides $\gamma_{12}=0.99$ eV/{\AA}$^2$ with a agreeable deviation of $0.08$  eV/{\AA}$^2$ (see Figs. (\ref{FigMOF1CC}e) and (\ref{FigMOF1CC}f)). Finally,
 the shearing (type (iv) deformation) provides $\gamma_{66}=0.49$  eV/{\AA}$^2$ and  $0.55$  eV/{\AA}$^2$, with a difference of $0.06$  eV/{\AA}$^2$ between the two methods (see Figs. (\ref{FigMOF1CC}g) and (\ref{FigMOF1CC}h)). Moreover, the relation $\gamma_{66}=\frac{1}{2}(\gamma_{11}-\gamma_{12})$ (because of the hexagonal symmetry) provides the values of $\gamma_{66}= 0.57$  eV/{\AA}$^2$ and  $0.48$ eV/{\AA}$^2$, demonstrating a trustworthy agreement. All these values of elastic constants are tabulated in Tab. (\ref{tab:MOF}). 

\subsection{MOF 2: Ni$_3$(C$_{6}$S$_{6}$)$_2$ }
Next, we take into account a metal-organic structure that is known to support Kagome bands.
The identification of Kagome bands in MOFs/COFs has triggered an 
extensive interest since 2013. The ground breaking work is the identification of nontrivial
topological states in an experimental 2D MOF sample,  Ni$_3$(C$_{6}$S$_{6}$)$_2$ lattice \cite{ZFW2},  which
comprises of three Ni ions being occupied at Kagome site (see  Fig. (\ref{FigMOF}b)). A monolayer of Ni$_3$(C$_{6}$S$_{6}$)$_2$  can be 
synthesized using an approach based on a bottom-up gas-liquid
interfacial reaction \cite{TK1}.   The band structure of Ni$_3$(C$_{6}$S$_{6}$)$_2$ 
shows one flat band above two Dirac bands near the Fermi level.

The DFT calculations are carried out  with a plane-wave cutoff of $480$ Ry on the $6 \times 6\times 1$ Monkhorst-Pack $k$-point mesh. As earlier, we consider a vacuum layer of $15$ {\AA}
thickness to avoid the  coupling between neighboring slabs and the atoms are relaxed until the forces are
smaller than $0.001$ Ry/Bohr. The optimized lattice constant is obtained  $14.64$ {\AA}, which is in good agreement with the
experimental value ($14-15$ {\AA}) \cite{TK2}. 

Now, we compute the elastic constants of MOF 2 in a similar manner as MOF 1 discussed in previous section, and obtain the values (see Tab. (\ref{tab:MOF}) and Fig. (\ref{FigMOF2CC})) of
 $\gamma_{11}=2.30$, $2.34$  eV/{\AA}$^2$,  $\gamma_{22}=2.35$,   $2.34$  eV/{\AA}$^2$,  $\gamma_{12}=0.95$, $0.93$  eV/{\AA}$^2$ and $\gamma_{66}=0.69$, $0.74$  eV/{\AA}$^2$. 
For all the elastic constants, the results from the stress-strain curve and total energy-strain curve  point to a solid agreement between two approaches. Here, it is to be noted that $\gamma_{11}$ and $\gamma_{22}$ have almost same values and hence confirming the relation $\gamma_{11}=\gamma_{22}$. Further, the relation $\gamma_{66}=\frac{1}{2}(\gamma_{11}-\gamma_{12})$ provides the values of $\gamma_{66}=0.68$, $0.71$  eV/{\AA}$^2$, presenting a solid accord.

\subsection{MOF 3: Cu$_3$(C$_6$N$_6$H$_6$)$_2$}
Finally, we take into consideration a metal-organic structure that serves as a prototype for combined honeycomb-Kagome (CHK) lattice systems. As the honeycomb and Kagome
lattices share the same unit cell, and the latter is the line graph of the former, the two can
be merged in one lattice, which is generally known as 
combined honeycomb-Kagome (CHK) lattice. 
Such a CHK lattice may be found in many MOFs and COFs, with the two sublattices being occupied by various organic ligands and/or metal ions, respectively \cite{WJ}.  Cu-hexaiminobenzene framework [Cu$_3$(HAB)$_2$, Cu$_3$(C$_6$N$_6$H$_6$)$_2$] is one of the prototype model systems of such a CHK lattice, where HAB ligands and Cu ions form the honeycomb and Kagome lattices, respectively, as illustrated in Fig. (\ref{FigMOF}c) \cite{WJ,WJ1}.

For the DFT computations, we consider a plane-wave cutoff of $480$ Ry and charge density cutoff of $80$ Ry on the $6 \times 6\times 1$ Monkhorst-Pack $k$-point mesh. Similar to earlier structures, with the relaxation of atoms until the forces are
smaller than $0.001$ Ry/Bohr and the vacuum space of $15$ {\AA} in the
$\hat{z}$-directions, the optimized lattice constant is found to have the value $14.33$ {\AA}.

Here, we compute the values of all elastic constants for MOF Cu$_3$(C$_6$N$_6$H$_6$)$_2$, where we find (see Fig. (\ref{FigMOF3CC}) and Tab. (\ref{tab:MOF})) $\gamma_{11}=1.51$, $1.52$ eV/{\AA}$^2$, $\gamma_{22}=1.65$, $1.55$  eV/{\AA}$^2$, $\gamma_{12}=0.46$, $0.35$  eV/{\AA}$^2$, and $\gamma_{66}=0.57$, $0.59$  eV/{\AA}$^2$.   Similar to previous structures, $\gamma_{11}$ and $\gamma_{22}$ should have the same values (because of the hexagonal symmetry), however they differ by $0.14$  eV/{\AA}$^2$ for the stress evaluation technique and $0.03$  eV/{\AA}$^2$ for the total energy evaluation method. This suggests that the value of $\gamma_{22}=1.65$  eV/{\AA}$^2$ is rather high in comparison. Moreover, using the relation $\gamma_{66}=\frac{1}{2}(\gamma_{11}-\gamma_{12})$, we obtain the values of $\gamma_{66}= 0.53$, $0.59$ eV/{\AA}$^2$, which indicates a substantial agreement as well.

Further, to have an idea about the hardness of these metal organic frameworks compared to a typical graphene sheet, we determine the elastic constants for a plane graphene sheet devoid of any hole, where we find $\gamma_{11}=20.95,20.92$ eV/{\AA}$^2$, $\gamma_{22}=20.95,20.94$ eV/{\AA}$^2$, 
$\gamma_{12}=4.95,4.99$ eV/{\AA}$^2$ and $\gamma_{66}=8.16,8.11$ eV/{\AA}$^2$, using both the approaches (stress and total energy evaluation). This finding indicates that the high porosity of these materials significantly reduces their hardness.
\subsection{COF 1: C$_9$N$_4$}
\begin{figure}[h!]
	\centering
	\includegraphics[width=0.7\textwidth]{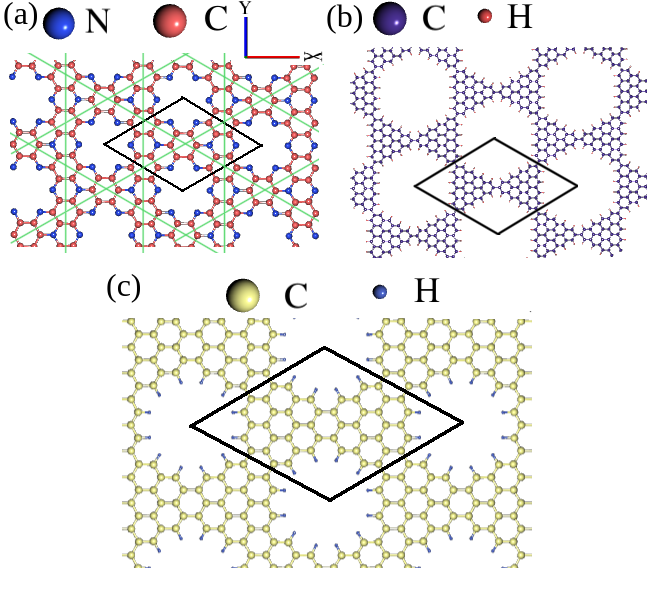}
	\caption{The atomic structures of (a) the C$_9$N$_4$ with the C rings (connected by green lines) forming a Kagome lattice, (b) and (c) representative COFs (with different sizes of nano holes) hosting enantiomorphic Kagome bands.  The black lines indicate the unit cells.
	}
	\label{FigCOF}
\end{figure}
\begin{table}[h!]
	\centering	  
	\begin{tabular}{| *{9}{c|} }	
		\hline
		& \multicolumn{2}{c|}{COF 1}
		& \multicolumn{2}{c|}{COF 2}
		& \multicolumn{2}{c|}{COF 3}
		\\
		\hline
		\thead {Elastic \\ Constants}  &   \thead {S-S\\(eV/{\AA}$^2$)}  &   \thead {TE-S\\ (eV/{\AA}$^2$)}  &    \thead {S-S\\(eV/{\AA}$^2$)}  &   \thead {TE-S\\ (eV/{\AA}$^2$)}  &    \thead {S-S\\eV/{\AA}$^2$)}  &   \thead {TE-S\\ (eV/{\AA}$^2$)}    \\
		\hline
		$\gamma_{11}$   &   13.20  &   13.02  &   2.99  &   3.31  &   8.58  &   8.52   \\
		\hline
		$\gamma_{22}$   &   13.07  &   12.88  &   3.54  &   3.42  &   8.58  &   8.51    \\
		\hline
		$\gamma_{12}$   &   2.79  &   2.93  &   1.40  &   1.34  &   2.29  &   2.42    \\
		\hline
		$\gamma_{66}$   &   4.98  &   4.99  &   1.04  &   0.96  &   3.07  &   3.12  \\
		\hline
	\end{tabular}\\
	\caption{Elastic constants of covalent-organic frameworks, where S-S  and TE-S stand for the calculations from stress-strain and total energy-strain curves, respectively.} 
	\label{tab:COF} 
\end{table}
We have so far talked about the metal-organic frameworks. Here, we concentrate on covalent-organic frameworks, where the first structure is a carbon nitride network.
Covalent compounds made of carbon nitride have recently emerged as a standout 2D material.
N appears to be a potential partner for combining with C to form an important set of 2D materials because of their close proximity to one another on the periodic table. Very recently, a group has reported the synthesization of  2D polyaniline with a C$_3$N
stoichiometry  using a direct
pyrolysis of hexaaminobenzene (HAB) trihydrochloride single
crystals \cite{JM}.  The C$_3$N monolayer is nothing more than an ordered, uniformly distributed version of N-substituted graphene. The experimentally achieved C-N compounds, such as polyaniline C$_3$N, 2D graphitic carbon nitride g-C$_3$N$_4$, and nitrogenated holey graphene C$_2$N, have a lot of potential for use in the environmental and energy sectors.
Here, we investigate a  novel type of covalent  carbon nitride network with a C$_9$N$_4$ \cite{HC} stoichiometry (see Fig. (\ref{FigCOF}a)). In contrast to common C-N compounds and covalent organic frameworks
(COFs) which are commonly insulating, C$_9$N$_4$ is ascertained to be a 2D nodal-line semimetal. As shown Fig. (\ref{FigCOF}a), the covalent C$_9$N$_4$
network depicts a periodicity of regular nano-holes, with the hexagonal C rings constituting a Kagome lattice (marked by green lines linking each center of the C rings) and the elements N constituting a honeycomb lattice.

For the DFT calculations, we consider a plane-wave cutoff of $480$ Ry on the $8 \times 8\times 1$ Monkhorst-Pack $k$-point mesh. Here also, we consider a empty space of at least $15$ $\AA $ in the
$\hat{z}$-directions to avoid the interaction between neighboring slabs.  The optimized lattice constant is obtained  $9.64$ $\AA$ with relaxation of atoms until the forces are
smaller than $0.001$ Ry/Bohr, indicating  a good agreement with the
value obtained in Ref. \cite{HC}.

Now, we compute the values of all elastic components for C$_9$N$_4$ structure with previously mentioned deformations from stress as well as total energy evaluation methods, where we find  (see Fig. (\ref{FigCOF1CC}) and Tab. (\ref{tab:COF}))  
 $\gamma_{11}=13.20$, $13.02$eV/{\AA}$^2$,
 $\gamma_{22}=13.07$,  $12.88$ eV/{\AA}$^2$, $\gamma_{12}=2.79$,  $2.93$ eV/{\AA}$^2$ and $\gamma_{66}=4.98$, $\gamma_{66}=4.99$ eV/{\AA}$^2$, while the relation $\gamma_{66}=\frac{1}{2}(\gamma_{11}-\gamma_{12})$ provides $\gamma_{66}=5.21$ $5.05$ eV/{\AA}$^2$, pointing a trustworthy agreement.  The elastic constants  $\gamma_{11}$ and $\gamma_{22}$ differ by $0.13$ eV/{\AA}$^2$ for stress evaluation
method and $0.14$ eV/{\AA}$^2$ for total energy evaluation
method, showing a reliable mismatch. For all elastic constants, the results suggest a reasonable agreement between two approaches.

\subsection{COF 2: A graphene nano-flake with enantiomorphic Kagome bands}

Now, we consider a  COF formed by graphene nano-flake as shown in Fig. (\ref{FigCOF}b) \cite{WJ1}. This COF hosts enantiomorphic Kagome bands \cite{YZ2} based on the ($s$,
$p_x$, $p_y$)-orbital honeycomb model. Around the Fermi level, the molecular orbitals (MOs) of the organic ligand  have the same symmetry and shape
as the atomic $s$, $p_x$ , and, $p_y$ orbitals, which lead
to the “super-atomic” orbital basis to form the
enantiomorphic Yin-Yang Kagome bands. Yin-Yang Kagome lattice is known to host 
non-equilibrium excited-state quantum Hall effect (EQHE) without any
intrinsic magnetization, which arises  between two enantiomorphic flat bands of opposite chirality from circularly polarized photoexcitation \cite{YZ2}.

DFT calculations are performed with a plane-wave cutoff of $480$ Ry on the $6 \times 6\times 1$ Monkhorst-Pack $k$-point mesh. Similar to previous structures, we consider a vacuum space of $15$ {\AA } in the
$\hat{z}$-directions to neglect the coupling between the images.  With the relaxation of atoms until the forces are
smaller than $0.001$ Ry/Bohr, the optimized lattice constant is obtained  $22.80$ {\AA}.

Now we calculate all the elastic constant values for the structure COF 2, where we obtain (see Fig. (\ref{FigCOF2CC}) and Tab. (\ref{tab:COF}))  $\gamma_{11}=2.99$, $3.31$ eV/{\AA}$^2$  (with a difference of $0.32$ eV$/\AA^2$ between two methods, which is comparatively a bit large discrepancy),  $\gamma_{22}=3.54$, $3.42$ eV/{\AA}$^2$, $\gamma_{12}=1.40$, $1.34$ eV/{\AA}$^2$ and $\gamma_{66}=1.04$, $0.96$ eV/{\AA}$^2$.   As mentioned earlier, because of the hexagonal symmetry of the structure,  $\gamma_{11}$ and $\gamma_{22}$ should have the same values, however our calculation shows they differ by $0.55$ eV/{\AA}$^2$ (for the stress evaluation technique) and $0.11$ eV/{\AA}$^2$ (for the total energy evaluation method), indicating comparably a greater numerical error for stress evaluation method. 
Further, the relation $\gamma_{66}=\frac{1}{2}(\gamma_{11}-\gamma_{12})$ provides $\gamma_{66}= 0.80$, $0.99$ eV/{\AA}$^2$, while the relation $\gamma_{66}=\frac{1}{2}(\gamma_{22}-\gamma_{12})$ (as $\gamma_{11}$ should have same value of $\gamma_{22}$ because of the lattice symmetry) provides $\gamma_{66}= 1.07$, $1.04$ eV/{\AA}$^2$. Thus, this comparison indicates the value of $\gamma_{11}=2.99$ eV/{\AA}$^2$ is a bit low and hence $\gamma_{66}= 0.80$ eV/{\AA}$^2$ is also a low value.

\subsection{COF 3}
Finally, we study another graphene nano-flake with a periodicity of regular hydrogen passivated nano-holes  with the hexagonal C rings constituting a Kagome lattice (linking each center of the C rings) as shown in Fig. (\ref{FigCOF}c). Similar to COF 2 Fig. (\ref{FigCOF}b), it also  shows the enantiomorphic Yin-Yang Kagome bands.

For the DFT calculations, as earlier, valence wave functions and
the deficit charge density are expanded in terms of the
plane-wave basis set with cutoff energies of $80$ Ry and $480$ Ry,
respectively, on the $6 \times 6\times 1$ Monkhorst-Pack $k$-point mesh. 
With the relaxation of atoms until the forces become
smaller than $0.001$ Ry/Bohr, we find that the optimized lattice constant is $28.12$ {\AA}. 

Now, we calculate the values of all elastic constants for COF 3, where we find (see Fig. (\ref{FigCOF3CC}) and Tab. (\ref{tab:COF}))    $\gamma_{11}=8.58$,  $8.52$ eV/{\AA}$^2$,  $\gamma_{22}=8.58$, $8.51$ eV/{\AA}$^2$,  $\gamma_{12}=2.29$,  $2.42$ eV/{\AA}$^2$ and
$\gamma_{66}=3.07$, $3.12$ eV/{\AA}$^2$.
 Thus, the
$\gamma_{11}$ and $\gamma_{22}$ values are exactly same for stress evaluation
method, whereas they differ by  only $0.01$ eV$\AA^2$ for total energy evaluation
method, confirming the relation $\gamma_{11}=\gamma_{22}$.  For all the elastic constants, the results from the stress-strain curve and total energy-strain curve  point to a solid agreement between two approaches.  Furthermore, using the relation $\gamma_{66}=\frac{1}{2}(\gamma_{11}-\gamma_{12})$, we obtain  $\gamma_{66}= 3.15$, $3.05$ eV/{\AA}$^2$, which as well indicates a solid agreement.

To comprehend how the values of various elastic constants are altered by the nanoholes, we compare our results  with a typical graphene sheet, which suggests that, depending on the size of the nanoholes, the hole reduces the values of the elastic constants. As the link between two nanoholes in COF $2$ is substantially narrower than that of COF $3$, we find  that the elastic constants of COF $3$ are larger than those of COF $2$.

\section{Conclusion}
\label{Sec4}
In summary, the mechanical properties of some 2D metal-organic frameworks (MOFs) and covalent-organic frameworks (COFs) with non trivial band dispersions are investigated using density functional theory (DFT).  In particular, we consider four types of deformation and find all the components of the elastic tensor  from the stress-strain and energy-strain curves. We apply the strains in the increments of $0.001$ in the region from $\eta =-0. 01$ to $+. 01$ (regime with linear dependency of
stress upon strain)  with the full geometrical optimization of the atomic structures and find the energy and stress at that value
of strain.  A nonlinear dependence can be found for higher
strain, however in this present study, we skip the nonlinear part. Ours results for all the structures obtained from two methods indicate a good agreement with each other. 
Because of the physical stability, unique structural properties, large surface to volume ratio, these porous materials can be potential candidates for technological applications and hence our findings may provide useful information to fabricate the MOFs/COFs based devices by lowering the number of experiments. 

\section{Acknowledgment}
This work is supported in part by JST CREST Grant No. JPMJCR20T3, Japan, and JSPS KAKENHI Grants No. JP21H05236, No. JP21H05232, No. JP20H01840, No. JP20H00127, and No. JP20K14415, Japan.

\vspace{2.0cm}\appendix{{\bf Appendix}}
\renewcommand\thefigure{A\arabic{figure}}  
\setcounter{figure}{0} 

\begin{figure} [h]
	\centering
	\includegraphics[width=1\textwidth]{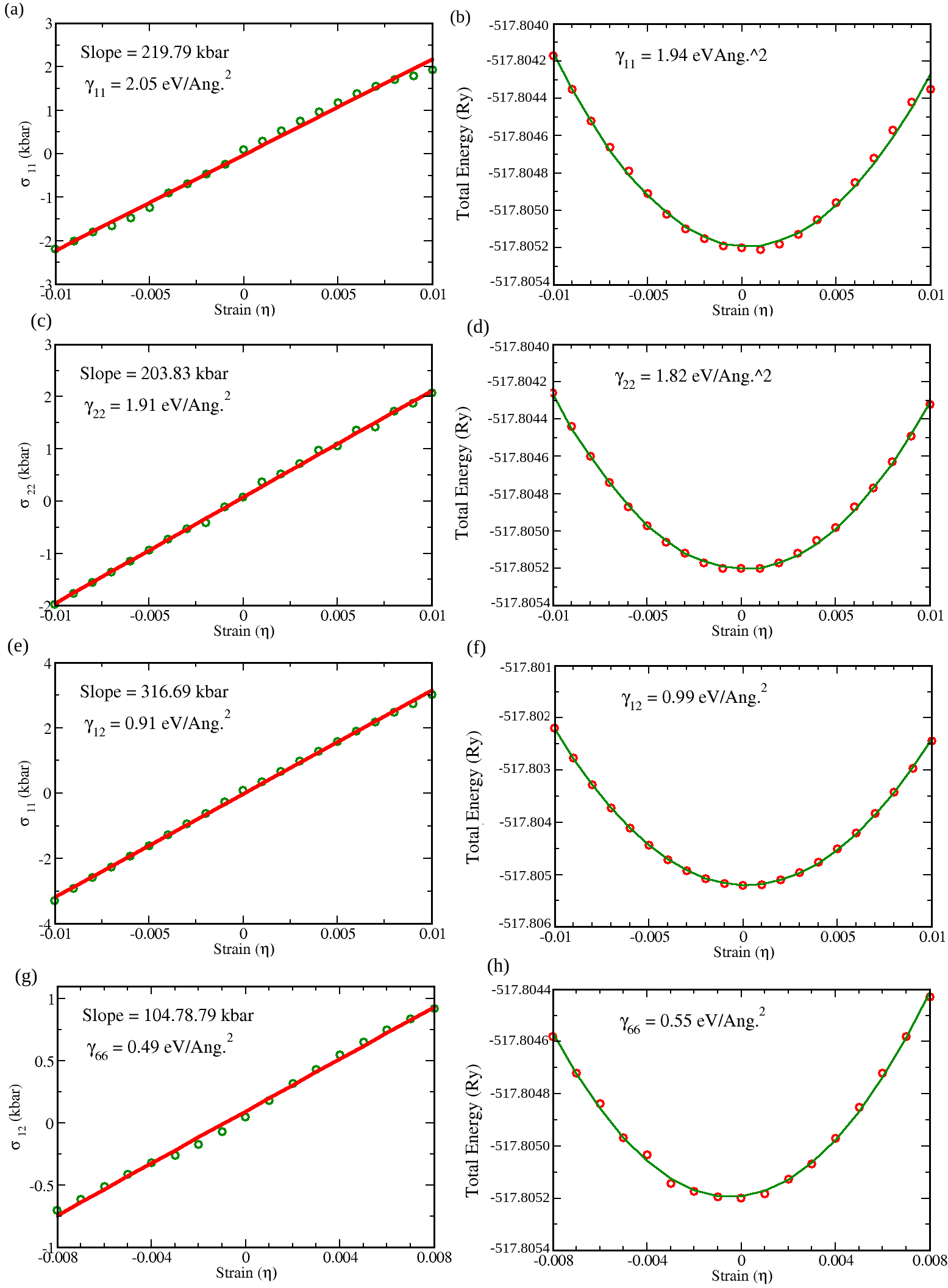}
	\caption{ MOF $1$: In$_2$(C$_6$H$_4$)$_3$. The left and right panels show the variation of  3D stress and the total energy as a function of strain $\eta$.  First, second, third and fourth rows describe the type (i), type (ii), type (iii) and type (iv) deformations, respectively. The green circles are obtained from DFT calculation, while the red solid line correspond to linear fitting for stress-strain curve and quadratic fitting for total energy-strain curve.}
	\label{FigMOF1CC}
\end{figure}
\begin{figure}
	\centering
	\includegraphics[width=1.0\textwidth]{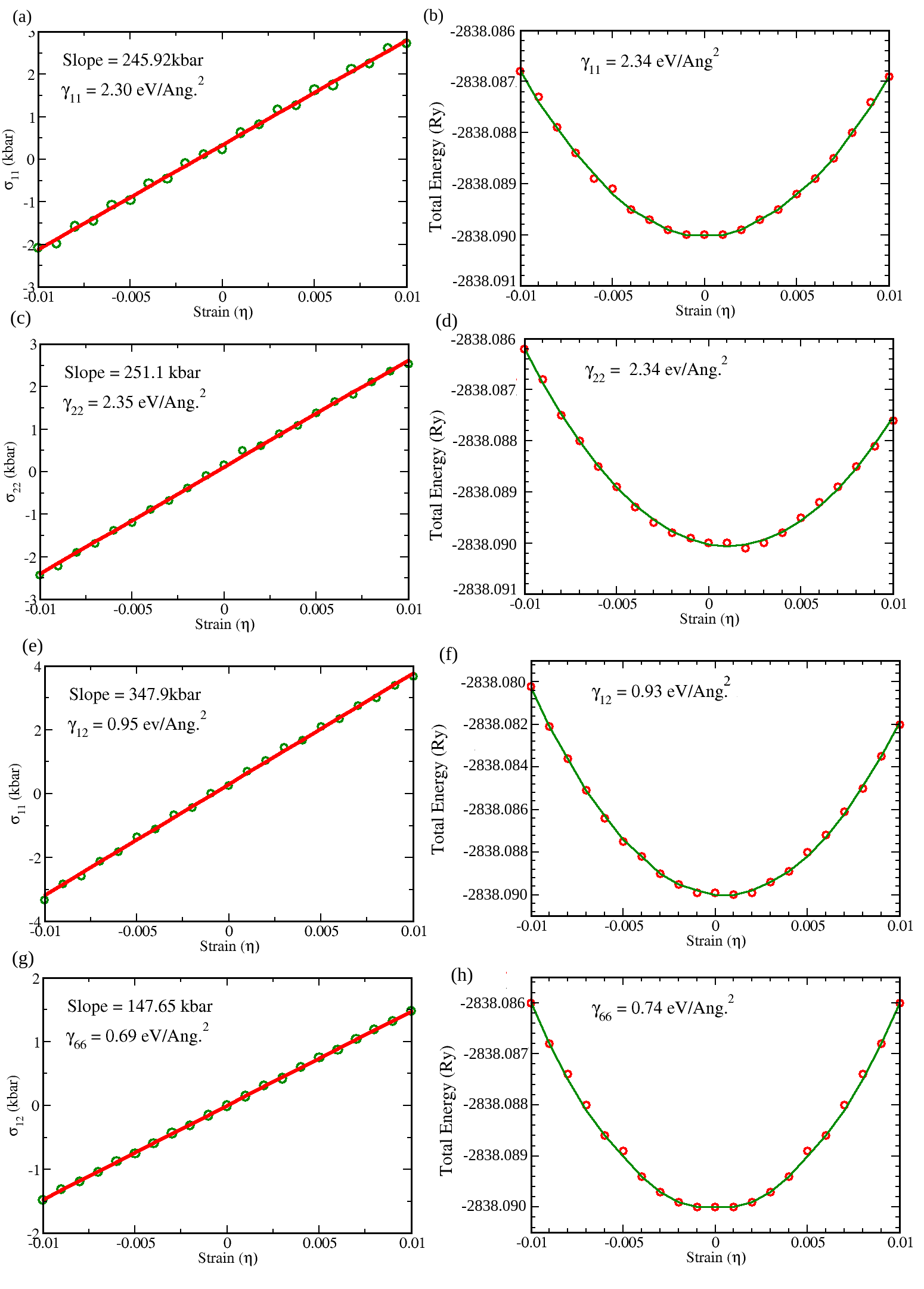}
	\caption{ MOF $2$: Ni$_3$(C$_{6}$S$_{6}$)$_2$. The Left and right panels show the variation of  3D stress and the total energy as a function of strain $\eta$.  First, second, third and fourth rows describe the type (i), type (ii), type (iii) and type (iv) deformations, respectively. The green circles are obtained from DFT calculation, while the red solid line correspond to linear fitting for stress-strain curve and quadratic fitting for total energy-strain curve. }
	\label{FigMOF2CC}
\end{figure}
\begin{figure}
	\centering
	\includegraphics[width=0.99\textwidth]{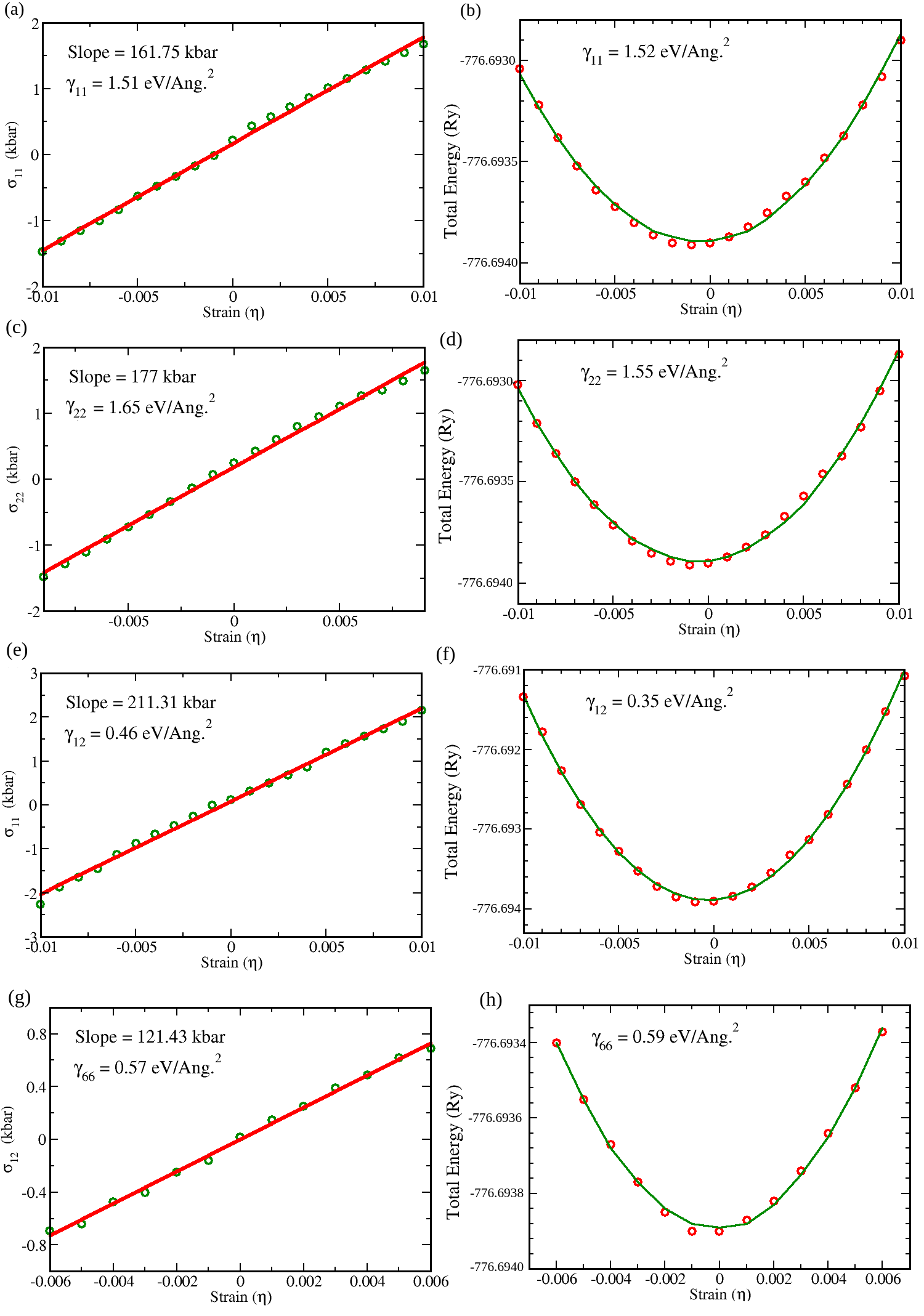}
	\caption{ MOF $3$: Cu$_3$(C$_6$N$_6$H$_6$)$_2$. The Left and right panels show the variation of  3D stress and the total energy as a function of strain $\eta$.  First, second, third and fourth rows describe the type (i), type (ii), type (iii) and type (iv) deformations, respectively. The green circles are obtained from DFT calculation, while the red solid line correspond to linear fitting for stress-strain curve and quadratic fitting for total energy-strain curve. }
	\label{FigMOF3CC}
\end{figure}
\begin{figure}
	\centering
	\includegraphics[width=0.99\textwidth]{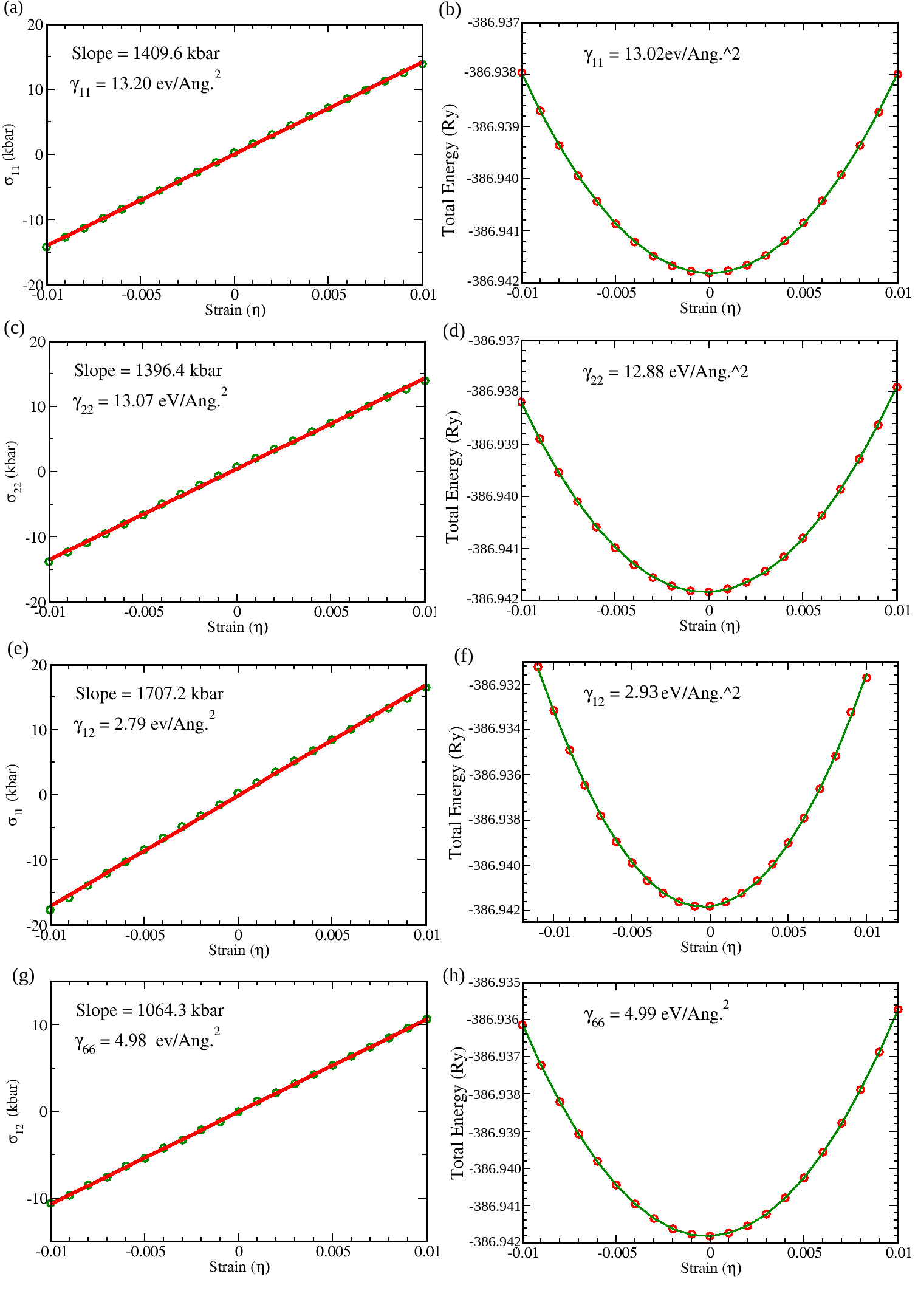}
	\caption{ COF $1$:  C$_9$N$_4$. The Left and right panels show the variation of  3D stress and the total energy as a function of strain $\eta$.  First, second, third and fourth rows describe the type (i), type (ii), type (iii) and type (iv) deformations, respectively. The green circles are obtained from DFT calculation, while the red solid line correspond to linear fitting for stress-strain curve and quadratic fitting for total energy-strain curve. }
	\label{FigCOF1CC}
\end{figure}
\begin{figure}
	\centering
	\includegraphics[width=1.0\textwidth]{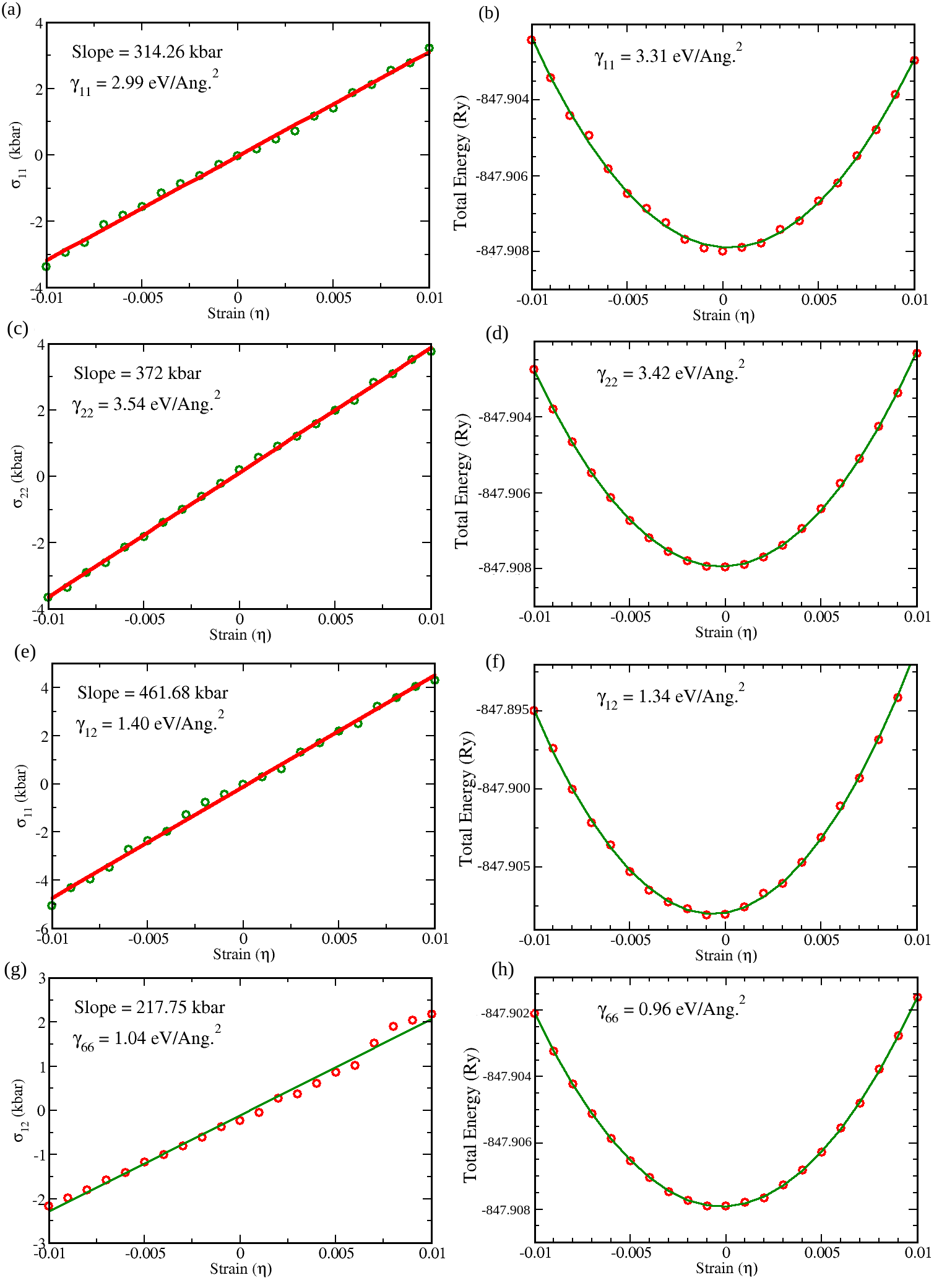}
	\caption{ COF $2$:  Representing COF hosting enantiomorphic Kagome bands.. The Left and right panels show the variation of  3D stress and the total energy as a function of strain $\eta$.  First, second, third and fourth rows describe the type (i), type (ii), type (iii) and type (iv) deformations, respectively. The green circles are obtained from DFT calculation, while the red solid line correspond to linear fitting for stress-strain curve and quadratic fitting for total energy-strain curve. }
	\label{FigCOF2CC}
\end{figure}
\begin{figure}
	\centering
	\includegraphics[width=1.0\textwidth]{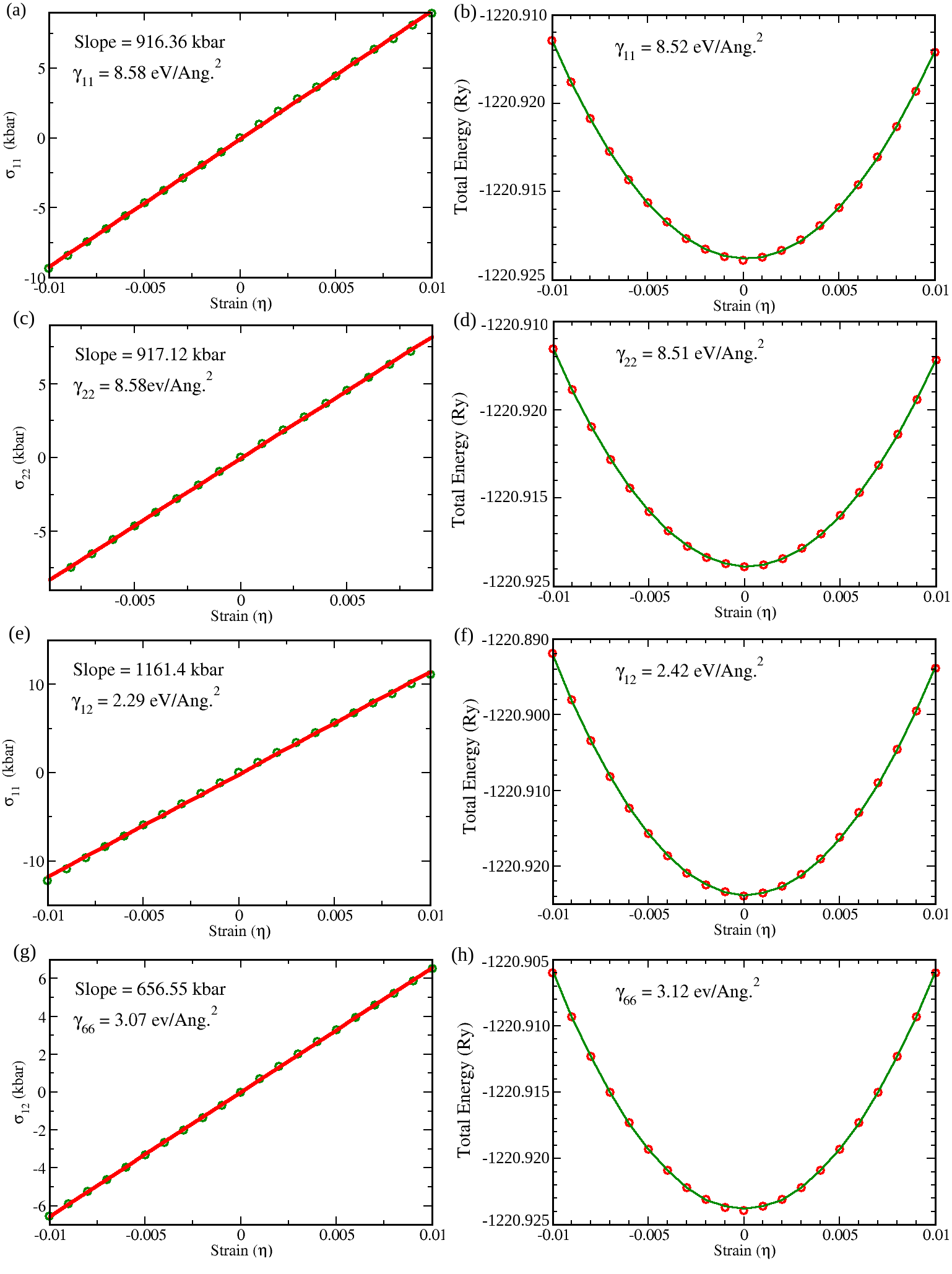}
	\caption{ COF $3$:  The Left and right panels show the variation of  3D stress and the total energy as a function of strain $\eta$.  First, second, third and fourth rows describe the type (i), type (ii), type (iii) and type (iv) deformations, respectively. The green circles are obtained from DFT calculation, while the red solid line correspond to linear fitting for stress-strain curve and quadratic fitting for total energy-strain curve. }
	\label{FigCOF3CC}
\end{figure}

This section contains all the plots for 3D stress and the total energy as a function of $\eta$ for the MOFs and COFs that we take into consideration in our study. Figures (\ref{FigMOF1CC}), (\ref{FigMOF2CC}) and (\ref{FigMOF3CC}) represent the MOFs 1,2 and 3, while Figs. (\ref{FigCOF1CC}), (\ref{FigCOF2CC}) and (\ref{FigCOF3CC}) represent the COFs 1,2,3. For all the plots, left panel depicts the variation of 3D stress and right panel shows the total energy as a function of $\eta$. The first, second, third, and fourth rows, respectively, describe the type $(i)$ $(ii)$, $(iii)$, and $(iv)$ deformations. 



\section{References}

\begin{thebibliography}{9}
	
	\bibitem{Nov1}
	K. S. Novoselov, A. K. Geim, S. V. Morozov, D. Jiang
	, M. I. Katsnelson, I. V. Grigorieva, S. V. Dubonos
	and A. A. Firsov, Nature {\bf 438}, 197 (2005).
	
	\bibitem{ADM}
	A. De Martino, L. DellAnna and R. Egger, Phys.
	Rev. Lett. {\bf 98}, 066802 (2007).
	
	\bibitem{CLK}
	C. L. Kane, E. J. Mele,
	Phys. Rev. Lett. {\bf 95}, 226801 (2005).
	
	\bibitem{BAB}
	A. B. Bernevig, T. L. Hughes, S. -C. Zhang, 
	Science (Washington, DC, U. S.) {\bf 314}, 1757 (2006).
	
	\bibitem{HT}
	H. Tasaki, 
	Phys. Rev. Lett. {\bf 69}, 1608 (1992).
	
	\bibitem{AM}
	A. Mielke, 
	J. Phys. A: Math. Gen. {\bf 25}, 4335 (1992).
	
	\bibitem{SM}
	S. Mukherjee, A. Spracklen, D. Choudhury, N. Goldman, 
	P. Ohberg, E. Andersson, R. R. Thomson, Phys. Rev. Lett. {\bf 114},
	245504 (2015).
	
	\bibitem{YZ1}
	Y. Zong, S. Xia, L. Tang, D. Song, Y. Hu, Y. Pei, J. Su, Y. Li, Z. Chen,   Opt. Express {\bf 24}, 8877 (2016).
	
	\bibitem{ZL1}
	Z. Liu, F. Liu, Y. S. Wu, 
	Chin. Phys. B {\bf 23}, 077308 (2014).
	
	
	\bibitem{ZFW1}
	Z. F.  Wang, K. Jin, F. Liu, WIREs Comput. Mol. Sci. {\bf 7},
	e1304  (2017).
	
	\bibitem{WJ}
	W. Jiang, F. Liu, F.  Materials and Energy; World
	Scientific Reference on Spin in Organics, pp. 201-224 (2018).
	
	\bibitem{AMS}
	 M. A. Springer, T. J. Liu, A. Kuc, T. Heine,  Chem. Soc. Rev. {\bf 49}, 2007 (2020).
	
	\bibitem{ZMH}
	Z. M. Hasan, C. L. Kane,
	Rev. Mod. Phys. {\bf 82}, 3045  (2010).
	
	\bibitem{XLQ}
	X. -L. Qi,  S. -C. Zhang, Rev. Mod. Phys. {\bf 83}, 1057 (2011).
	
	\bibitem{XF}
	X. Feng, X. Ding, D. Jiang, 
	Chem. Soc. Rev. {\bf 41}, 6010 (2012).
	
	\bibitem{HCJZ}
	H. C. J. Zhou, S. Kitagawa, 
	Chem. Soc. Rev. {\bf 43}, 5415 (2014).	
	
	\bibitem{AC}
	A. Croy, D. Midtvedt, A. Isacsson, J. M. Kinaret, Phys. Rev. B {\bf 86}, 235435 (2012).
	
	\bibitem{BR}
	B. Radisavljevic, A. Radenovic, J. Brivio, V. Giacometti, A. Kis,
	Nat. Nanotechnol. {\bf 6},  147 (2011).
	
	\bibitem{PL}
	P. Li, Z. You, T. Cui,
	Appl. Phys. Lett. {\bf 101}, 093111 (2012).
	
	\bibitem{HB} H. Bu, Y. Chen, M. Zou, H. Yi, K. Bi, Z. Ni,
	Phys. Lett.  A {\bf 373}, 3359 (2009).
	
	
	\bibitem{HZ} H. Zhao, K. Min, N. R. Aluru,
	Nano Lett. {\bf 9}  3012 (2009).
	
	\bibitem{CR} C. D. Reddy, A. Ramasubramaniam, V. B. Shenoy, Y. W. Zhang,
	Appl. Phys. Lett. {\bf 94}, 101904 (2009).
	
	\bibitem{RF} R. Faccio, P. A. Denis, H. Pardo, C. Goyenola, A. W. Mombru, J. Phys.: Condens. Matter {\bf 21}, 285304 (2009).
	
	
	\bibitem{AVO} A.V. Orlov, and I. A. Ovid'ko,
	Rev. Adv. Mater. Sci. {\bf 40}, 249 (2015).
	
	\bibitem{JFN} J. F. Nye,  Physical Properties of Crystals. London: Oxford University Press, (1959).
	
	\bibitem{MN}
	M Niederreiter, Determining elastic constants of
	graphene using density functional theory, University Of Graz (2018).
	
	\bibitem{PG1}
	P. Giannozzi, S. Baroni, N. Bonini, M. Calandra, R. Car,
	C. Cavazzoni, D. Ceresoli, G. L. Chiarotti, M. Cococcioni,
	I. Dabo et al.,
	J. Phys.: Condens. Matter {\bf 21}, 395502 (2009).
	
	\bibitem{PG2}
	P. Giannozzi, O. Andreussi, T. Brumme, O. Bunau, M. B.
	Nardelli, M. Calandra, R. Car, C. Cavazzoni, D. Ceresoli, M.
	Cococcioni et al., 
	J. Phys.: Condens. Matter {\bf 29}, 465901 (2017).
	
	\bibitem{ZL2}
	Z. Liu, Z. F. Wang, J. W. Mei, Y. S. Wu, F Liu,
	Phys. Rev. Lett. {\bf 110}, 106804 (2013).
	
	\bibitem{ZFW2} Z. F. Wang, N. Su, F. Liu, 
	Nano Lett. {\bf 13}, 2842, (2013).
	
	\bibitem{TK1}
	T. Kambe, R. Sakamoto, T. Kusamoto, T. Pal, N. Fukui, 
	K. Hoshiko, T. Shimojima, Z. Wang, T. Hirahara, K. Ishizaka, 
	S. Hasegawa, F.  Liu, H. Nishihara,  J. Am.
	Chem. Soc. {\bf 136}, 14357, (2014).
	
	
	\bibitem{TK2}
	T. Kambe, R. Sakamoto, K. Hoshiko, K. Takada, M. Miyachi,
	J. -H. Ryu, S. Sasaki, J. Kim, K. Nakazato, M. Takata, H. Nishihara, 
	J. Am. Chem. Soc.  {\bf 135}, 2462 (2013).
	
		\bibitem{WJ1}
	W. Jiang,  X. Ni, § and F. Liu,
	Acc. Chem. Res. {\bf 54}, 416 (2021).
	
	
	
	\bibitem{JM}
	J. Mahmood, E. K. Lee, M. Jung, D. Shin, H. J. Choi, J. M. Seo,
	S. M. Jung, D. Kim, F. Li, M. S. Lah, N. Park, H. J. Shin,
	J. H. Oh and J. B. Baek,
	Proc. Natl. Acad. Sci. U. S. A
	{\bf 113}, 7414 (2016).
	
	\bibitem{HC}
	H. Chen, S. Zhang, W. Jiang, C. Zhang, H. Guo, Z. Liu, Z. Wang, F. Liu, X. Niu, 
	J. Mater. Chem. A {\bf 6}, 11252 (2018).
	

	
	\bibitem{YZ2}
	Y. Zhou, G. Sethi, H. Liu, F. Liu, F.  arXiv
	(Materials Science): 1908.03689, ver. 2. https://arxiv.
	org/abs/1908.03689, (2019).	
	
\end{thebibliography}
\end{document}